\documentclass[12pt,preprint]{aastex}

\shorttitle{Influence of temperature fluctuations}
\shortauthors{Silant'ev et al.}

\begin{document}

\title {Influence of temperature fluctuations on \\continuum spectra of
cosmic objects}

\author{ N.~A.~Silant'ev\altaffilmark{1}, G.~A.~Alekseeva\altaffilmark{1} \& V.~V.~Novikov\altaffilmark{1}}

\affil{Main (Pulkovo) Astronomical Observatory of Russian
 Academy of Sciences,\\ Pulkovskoe shosse 65, St.- Petersburg, 196140, Russia}
\email{ nsilant@bk.ru}

\begin{abstract}
The presence of convective and turbulent motions, and the evolution of magnetic
fields give rise to existence of temperature fluctuations in stellar atmospheres, active galactic
nuclei and other cosmic objects. We observe the time and surface averaged radiation fluxes from these objects.
These fluxes depend on both the mean temperature and averaged temperature fluctuations.
The usual photosphere models do not
take into account the temperature fluctuations and use only the distribution of the mean
temperature into surface layers of stars. We investigate how the temperature fluctuations
change the spectra in continuum assuming that the degree of fluctuations (the ratio
of mean temperature fluctuation to the mean temperature) is small. We suggest the procedure
of calculation of continuum spectra, which takes into account the temperature fluctuations. As a first
step one uses the usual model of a photosphere without fluctuations.
The observed spectrum is presented as a part depending on mean temperature and the additional part
proportional to quadratic value of fluctuation degree. It is shown that for some forms of absorption factor
the additional part in Wien's region of spectrum can be evaluated directly from
observed spectrum. This part depends on the first and second wavelength derivatives, which can be calculated
numerically from the observed spectrum. Our estimates show that the temperature
dependence of absorption factors is very important by calculation of continuum spectra corrections.
As the examples we present the estimates for a few stars from
Pulkovo spectrophotometric catalog and for the Sun. The influence of temperature fluctuations on color indices
of observed cosmic objects is also investigated.
\end{abstract}


\newpage

\section{Introduction}

From investigation of the Sun \citep[see] []{sti91} we know that the photosphere has chaotic changes of the temperature.
They are due to convective and turbulent motions, and the existence of inhomogeneous magnetic fields,
 which
are the specific examples of stochastic media. Recall that stochastic behavior of gas motions arises due to
many types of instabilities occurring in these media. Most important is the convective instability which results in
the star cells of rising and descending motions of hot and cold gas, correspondingly.
For the Sun the ratio of mean temperature fluctuations $T'$ to the mean temperature $\langle T\rangle \equiv T_0$
is equal to $\eta =\sqrt{\langle T'^2\rangle}/T_0 \approx 0.03$. It appears this estimation is minimal because
the solar spots are not taken into account.
No doubt, the temperature fluctuations exist in many stars, not only in the Sun. The Sun
is the only star which can be observed detailed in every point of the surface. Further we restrict ourselves by
consideration of point-like cosmic objects.
Apparently the most high fluctuations exist in stars with the profound convective zone and in active galactic
nuclei.

In stochastic media all the values - the temperature $T=\langle T\rangle+T'$, the radiation intensity
$I=\langle I\rangle+I'$, the absorbtion coefficient
$\alpha_{\lambda}=\langle\alpha_{\lambda}\rangle+\alpha_{\lambda}'$ and so on, are stochastic, i.e. are characterized
by its mean values, for example, $\langle T\rangle\equiv T_0$ and by fluctuations $T'$.
The mean value of fluctuations are equal to zero (for example, $\langle T'\rangle =0$).
 The mean values are determined by the average
over ensemble of realizations during the time of observation. The chaotic events in space and time with
characteristic life-time greater
than the observation time are taken into account in the mean values  \citep[see, in more detail,] []{lev97}.
From the stars and other point-like cosmic objects we observe the mean radiative
flux {\bf $\langle H\rangle $}. We stress that the influence of temperature fluctuations on the spectra is pure
statistical effect, depending on the specific ensemble of temperature realizations in space and time, and on the
time of observation.
It is known that the averaged spectra of stars, even neighboring
in spectral classes, frequently differ one another \citep[see catalog] []{ale96}. The possible reasons may
be different, and the influence of temperature fluctuations, without doubt, is one of the most important between them.

In present paper we investigate the influence of temperature fluctuations on the continuum spectra of various
cosmic objects. The preliminary consideration of this subject is given in the paper \citep{sil08}.
Their paper deals only with investigation of effective thermal sources in stochastic media
$S_{\lambda}=\langle \alpha_{\lambda}(T)B_{\lambda}(T)\rangle $ as compared with the usual sources,
$\alpha_{\lambda}(T_0)B_{\lambda}(T_0)$, taken without the influence of temperature fluctuations.

Here $B_{\lambda}(T)$ is the intensity of Planck's radiation. It was shown that the source $S_{\lambda}$ can
fairly strong differ from $\alpha_{\lambda}(T_0)B_{\lambda}(T_0)$, especially at high value of parameter
$h\nu/kT_0$. The value $S_{\lambda}$ can be both higher than $\alpha_{\lambda}(T_0)B_{\lambda}(T_0)$ and lower one,
depending on specific form of absorbtion coefficient $\alpha_{\lambda}(T)$. Below we will obtain the formula for
corrections to observed radiation flux, which arises due to existence of temperature fluctuations.

Before start a detail consideration let us show in a simple example of averaging over two temperature realizations
($T=T_0+T'$ and $T=T_0-T'$), how arises this statistical effect. The average value of Planck's function (in
the Wien limit) is equal to:

\[
\langle B_{\lambda}(T)\rangle =\frac{1}{2}\,\left(\frac{2hc^2}{\lambda^5}\right)
\left\{\exp{\left[-\left(\frac{h\nu}{k(T_0+T')}\right)\right]}+\right.
\]

\[
\left. \exp{\left[-\left(\frac{h\nu}{k(T_0-T')}\right)\right]}\right\}\simeq
\]

\[
B_{\lambda}(T_0)\cosh{\left(\frac{h\nu}{kT_0}\cdot\frac{T'}{T_0}\right)}\ge B_{\lambda}(T_0) .
\]

\noindent Here we used the condition $T'/T_0\ll 1$.

It is seen that the observing averaged value $\langle B_{\lambda}(T)\rangle $ is higher than $B_{\lambda}(T_0)$.
At $T'/T_0=0.05$ and $h\nu/kT_0=10$ this growing consists of 13\%. At the Rayleigh limit ($h\nu/kT\ll 1$)
the Planck function $B_{\lambda}\sim T$ and  $\langle B_{\lambda}\rangle =B_{\lambda}(T_0)$. For intermediate
cases one has $\langle B_{\lambda}(T)\rangle > B_{\lambda}(T_0)$.

The analogous calculations demonstrate that the mean value of absorbtion factor
$\langle \alpha_{\lambda}(T)\rangle$ can be both higher than $\alpha_{\lambda}(T_0)$ and lower than this value,
depending on the specific form of $\alpha_{\lambda}(T)$.
 Joint consideration of Planck function and absorbtion coefficient can give
rise both to positive and negative  additional contribution to the usual spectra, depending on the mean temperature
$T_0$.

Below we investigate the problem how the temperature fluctuations change
the continuum spectra of cosmic objects in detail. For a number of stars and wavelengths we derived numerically
the corresponding corrections to the spectra, depending on the degree of temperature fluctuations. For the cases, when
the main contribution is due to bound-free transitions in negative hydrogen ions, we suggest the new techique to
calculate the corrections to the spectra directly from the shape of observing spectra, without using the results of
any photosphere models. Note that we consider a degree of temperature fluctuations $\eta $ as a given parameter.
The inverse problem - the estimation of $\eta$ from observed spectra is not considered in our paper. We hope investigate
this difficult problem in the future.

\section{Basic formulas}

The radiation flux in continuum $H^{(0)}_{\lambda}$ is related with the Planck function
$B_{\lambda}$ as follows \citep[see] []{sob69}:

\[
 H^{(0)}_{\lambda}=2\pi\int^1_0d\mu\mu I_{\lambda}(0,\mu)=2\pi\int^{\infty}_0d\tau_{\lambda}\int^1_0
 d\mu e^{-\tau_{\lambda}/\mu}
 B_{\lambda}(T_0(\tau_{\lambda}))=
\]
\begin{equation}
2\pi\int^{\infty}_0d\tau_{\lambda}E_2(\tau_{\lambda})B_{\lambda}(T_0(\tau_{\lambda})).
\label{1}
\end{equation}
\noindent Recall the notion of functions $E_n(\tau)$ \citep[see] []{sob69,gra76}:

\begin{equation}
 E_n(\tau)=\int^1_0d\mu\mu^{n-2}e^{-\tau/\mu}=\int^{\infty}_1\frac{dx}{x^n}e^{-\tau x}.
\label{2}
\end{equation}
\noindent The functions $E_n(\tau)$ obey the relations:

\[
 nE_{n+1}(\tau)=e^{-\tau}-\tau E_n(\tau),
\]
\begin{equation}
 \frac{dE_n(\tau)}{d\tau}=-E_{n-1}(\tau).
\label{3}
\end{equation}
\noindent At high $\tau $ one exists $E_n(\tau)\to \exp(-\tau)/\tau$. The figures $E_1(\tau), E_2(\tau)$
and $E_3(\tau)$, for example, are given in \citep{gra76}.

In Eq.(1) the mean temperature $T_0$ is considered as known function of the optical depth $\tau_{\lambda}$.
To find the relation between the mean temperature $T_0$ and optical depth $\tau_{\lambda}$ is the main task
at construction of photosphere models.
Formula (1) does not take into account the influence of temperature fluctuations.

In the paper \citep{sil05} the radiative transfer equation for mean intensity
 $\langle I_{\lambda}\rangle\equiv I^{(0)}_{\lambda}(\tau_{\lambda}, \mu)$ was derived. This equation has
the effective source $S_{\lambda}$, which was investigated  by \citep{sil08}. The sign
$\langle \,\rangle$ in radiative transfer equation (4) means the averaging over ensemble of various realizations
along the wave path, i.e. one takes into account the local temperature fluctuations.

In LTE approximation the transfer equation for $I^{(0)}_{\lambda}(\tau_{\lambda},\mu)$
acquires the form:

\begin{equation}
 ({\bf n}\nabla)I^{(0)}_{\lambda}=-\langle\alpha_{\lambda}\rangle I^{(0)}_{\lambda}+S_{\lambda}(T_0,\eta),
\label{4}
\end{equation}

\begin{equation}
 S_{\lambda}(T_0, \eta)=\langle \alpha_{\lambda}(T)B_{\lambda}(T)\rangle.
\label{5}
\end{equation}
\noindent As a result of averaging, all the values acquire the dependence on the degree of temperature fluctuations:

\begin{equation}
\eta=\frac{\sqrt{\langle T'^2\rangle}}{T_0} .
\label{6}
\end{equation}

The fluctuations $T'$ have the local character. Thus, in the case of the Sun they describes
the difference between hot and cold parts of granula.
The transfer equation (4) is stationary and differs from the usual transfer equation only by the feature that
the temperature fluctuations are taken into account in  absorbtion coefficient
$\alpha_{\lambda}(T)=\alpha_{\lambda}(T_0+T')$. Usually in photosphere models one assumes that all physical
values are distributed sperically symmetric. We assume the same for temperature fluctuations, i.e. the degree of
fluctuations $\eta$ is independent of the place in visual semisphere of a star.

 According to \citep{sil05}, more exact expression takes into
account the  correlation length  $R$ of temperature fluctuations, i.e. instead of term
$\langle \alpha_{\lambda}\rangle$ one uses the expression
 $\langle \alpha_{\lambda}\rangle\cdot (1+\langle (\alpha')^2 \rangle/(\langle \alpha \rangle)^2 \cdot \tau_{\lambda})$.
Here $\tau_{\lambda}=\langle \alpha_{\lambda} \rangle\cdot R$ is the mean optical depth of characteristic length
of a turbulence $R$.
Usually one exists $\tau_{\lambda}<<1$, i.e. the additional term is of the third degree of smallness and can be
omitted.
Such case takes place for solar granulations with  $R\simeq 1000$ km in the region of wavelength, where the basic
absorbtion is determined by negative hydrogen ion.

From Eqs. (4) and (5) one follows the expression for observed radiation flux $\langle H_{\lambda}\rangle $, which takes
into consideration the influence of temperature fluctuations:

\[
 \langle H_{\lambda}(\eta)\rangle =2\pi\int^1_0d\mu \mu I^{(0)}_{\lambda}(0,\mu)=
\]

\begin{equation}
 2\pi\int^{\infty}_0d\langle\tau_{\lambda}\rangle E_2(\langle \tau_{\lambda}\rangle )
 \frac{\langle\alpha_{\lambda}(T)B_{\lambda}(T)\rangle }{\langle\alpha_{\lambda}(T)\rangle}.
\label{7}
\end{equation}

Mean values $\langle\alpha_{\lambda}(T)\rangle $ are derived by the gaussian formula (see Silant'ev \& Alekseeva 2008):

\begin{equation}
 \langle\alpha_{\lambda}(T)\rangle=\frac{1}{\sqrt{2\pi}\eta }\int^{\infty}_{-\infty}dx
 \exp{\left(-\frac{x^2}{2\eta^2}\right)}\alpha_{\lambda}(T_0\cdot(1+x)).
\label{8}
\end{equation}
\noindent This expression is valid at $\eta\le 0.2$.

When the fluctuations are absent, $\eta=0$, formula (7) reverts to usual expression (1) for $H^{(0)}_{\lambda}$,
where $d\tau_{\lambda}=\alpha_{\lambda}(T_0)dz$
and $T_0\equiv T_0(\tau_{\lambda})$. For brevity we will use simply $T_0$. Recall that
the radiation flux without consideration of fluctuations we denote as $H_{\lambda}^{(0)}$.
The value $\langle \alpha_{\lambda}(T)B_{\lambda}(T)\rangle$ is calculated from Eq.(8) analogously.

Our next task is to relate the observing flux $\langle H_{\lambda}(\eta)\rangle $ with the
$H^{(0)}_{\lambda}$, i.e. to show how the small temperature fluctuations change the continuum spectra.
As an initial approximation we take the expression $H^{(0)}_{\lambda}$, and then we will seek the corrections
to this spectrum proportional to $\eta^2$. In section 4 we will consider this problem in Wien's region of spectrum,
and for such temperatures and wavelengths where the main absorbtion is due to bound-free transitions of
outer electron in negative hydrogen ion. In this case the flux $\langle H\rangle$  can be found by
numerical differentiation of observed spectra over wavelength $\lambda$.

The values of fluctuation degree $\eta $ are determined by physical conditions in every specific object.
We assume $\eta $ as a known parameter and derive the formula for estimates of spectra corrections due to temperature
fluctuations for a number of objects, where absorbtion coefficient deals with bound-free transitions in the
negative hydrogen ions.
Of course, it would be very interesting and important to derive the technique
how to estimate the parameter $\eta$ from the analysis of observed spectra. It appears this complex task may
be resolved by statistical comparison of objects with very close physical conditions. In this paper
we do not consider this {\bf difficult} problem.

\section{Relation of flux $\langle H_{\lambda}(\eta)\rangle $ with $H_{\lambda}^{(0)}$}

Assuming that the temperature fluctuations are low ($\eta^2 \ll 1$), let us use in Eq.(8) the expansion series:

\begin{equation}
 \alpha_{\lambda}(T_0\cdot(1+x))= \alpha_{\lambda}(T_0)(1+b_{\lambda}(T_0)x+a_{\lambda}(T_0)x^2+...),
\label{9}
\end{equation}
\noindent and take into mind that $\langle x\rangle=0$ and $\langle x^2\rangle=\eta^2$.
Dimensionless parameters $b_{\lambda}(T_0)$ and $a_{\lambda}(T_0)$ have the forms:

\[
 b_{\lambda}(T_0)=\frac{T_0}{\alpha_{\lambda}(T_0)}\,\frac{\partial \alpha_{\lambda}(T_0)}{\partial T_0},
\]
\begin{equation}
 a_{\lambda}(T_0)=\frac{T_0^2}{2\alpha_{\lambda}(T_0)}\,\frac{\partial^2 \alpha_{\lambda}(T_0)}{\partial T_0^2}.
\label{10}
\end{equation}
\noindent Thus, the mean absorbtion factor acquires the form:

\begin{equation}
 \langle \alpha_{\lambda}\rangle =\alpha_{\lambda}(T_0)(1+a_{\lambda}(T_0)\,\eta^2).
\label{11}
\end{equation}

The optical depth $\langle \tau_{\lambda}\rangle$ in expression (7) is calculated according to formula:

\begin{equation}
 \langle\tau_{\lambda}\rangle=\int^z_0dz\,\langle\alpha_{\lambda}(z)\rangle =
 \tau_{\lambda}+\eta^2\int^{\tau_{\lambda}}_0d\tau'_{\lambda} \,a_{\lambda}(T'_0),
\label{12}
\end{equation}
\noindent where we use the same notion  $d\tau_{\lambda}=\alpha_{\lambda}(T_0(z))dz$, as in calculation of the flux
$H^{(0)}_{\lambda}$ (see Eq. (1)).

Calculating by analogy the value $\langle\alpha_{\lambda}(T)B_{\lambda}(T)\rangle$, we
obtain from Eq. (7) the following expression for observed flux $\langle H_{\lambda}(\eta)\rangle$:

\[
 \langle H_{\lambda}(\eta)\rangle=H^{(0)}_{\lambda}+
\]
\begin{equation}
2\pi \eta^2\int^{\infty}_0d\tau_{\lambda}\,E_2(\tau_{\lambda})
 \left[b_{\lambda}(T_0)\,T_0\,\frac{\partial B_{\lambda}(T_0)}{\partial T_0}+\frac{1}{2}\,T^2_0\,
 \frac{\partial^2 B_{\lambda}(T_0)}{\partial T_0^2}\right]+ A_{\lambda},
\label{13}
\end{equation}

\begin{equation}
 A_{\lambda}=2\pi \eta^2\int^{\infty}_0d\tau_{\lambda}\,B_{\lambda}(T_0)\left[E_2(\tau_{\lambda})\,a_{\lambda}(T_0)-
 E_1(\tau_{\lambda})\int^{\tau_{\lambda}}_0d\tau_{\lambda}' \,a_{\lambda}(T_0')\right].
\label{14}
\end{equation}
\noindent The value $T_0'\equiv T_0(\tau_{\lambda}')$. The negative term with $a_{\lambda}(T_0')$
in Eq.(14) arises from the expansion of the exponent $\exp(-\langle\tau_{\lambda}\rangle)$
(see formula (12)) in power series with parameter $\eta^2$.

Using the known photosphere models, i.e. the dependence between $T_0$ and optical depth $\tau_{\lambda}$, one can
calculate the right part of Eq.(13) and obtain the flux $\langle H_{\lambda}(\eta)\rangle$, corresponding to
level of temperature fluctuation $\eta$. The second and third terms in Eq. (13) characterize
correction  to model spectrum $H^{(0)}_{\lambda}$ due to temperature fluctuations.
In principle, the condition of the best
coincidence of expression $\langle H_{\lambda}(\eta)\rangle$  with observed flux allows us to estimate parameter
$\eta $ in the atmosphere of a cosmic object. The influence of fluctuations is most strong in the region of
short waves. It appears we have to became the best coincidence just in this wave region. Taking different values
for parameter $\eta$, one can obtain more satisfactory coincidence of flux $\langle H_{\lambda}(\eta)\rangle$ with
the observed spectrum than that for model case $H^{(0)}_{\lambda}$.

The authors are not
specialists in the model construction activity. Therefore we restrict ourselves by the estimate of difference
$(\langle H_{\lambda}(\eta)\rangle -H^{(0)}_{\lambda})/H^{(0)}_{\lambda}$ only for a number of Sun-like stars and
the Sun for wavelengths where main contribution to opacity gives the bound-free electron transitions
in negative hydrogen ions $H^{-}$. These estimates follow directly from observed spectra \citep[see catalog] []{ale96}.
For these wavelengths we can substitute the temperature derivatives by numerical derivatives over wavelengths
in observed spectra (the results are presented in Tables 2 and 3).

First we yield this substitution for Plank's function $B_{\lambda}(T_0)$. Then such
substitution will be made for some types of absorbtion coefficients. Note that not all types of $\alpha(T_0)$
allow us such substitution. Finally, we will consider the spectrum in Wien's region, simultaneously using
the specific form of bound-free transitions in negative hydrogen ion $H^{-}$ for those cases, where this absorbtion
plays the dominant role.

\subsection{Substitution temperature derivatives by wavelength derivatives}

The Planck function has the following form:

\[
B_{\lambda}(T)=\frac{2hc^2}{\lambda^5}f(\lambda T), \,\,\,\,\,\, f(\lambda T)=\frac{1}{e^{g_0/\lambda T}-1},
\]
\begin{equation}
\frac{g_0}{\lambda T}\equiv\frac{h\nu}{kT}=\frac{14388}{\lambda T}.
\label{15}
\end{equation}
\noindent Here and what follows the wavelengths always are taken in microns.
The function $f(\lambda T)$, depending on the product $\lambda $ and $T$, obeys the relations:

\begin{equation}
 \lambda\,\frac{\partial f}{\partial \lambda}=T\,\frac{\partial f}{\partial T},\,\,\,\,\,
 \lambda^2\,\frac{\partial^2 f}{\partial \lambda^2}=T^2\,\frac{\partial^2 f}{\partial T^2}.
\label{16}
\end{equation}

These relations allow us to substitute the derivatives over temperature by the derivatives over wavelength.
The corresponding formulas are the follows:

\[
\frac{\partial B_{\lambda}}{\partial T}=\frac{1}{T}\left(5B_{\lambda}+\lambda \,\frac{\partial B_{\lambda}}
{\partial \lambda}\right),
\]
\begin{equation}
 \frac{\partial^2 B_{\lambda}}{\partial T^2}=\frac{1}{T^2}\left(20B_{\lambda}+10\lambda \,\frac{\partial B_{\lambda}}
{\partial \lambda }+\lambda^2 \,\frac{\partial^2 B_{\lambda}}{\partial \lambda^2}\right) .
\label{17}
\end{equation}

Absorbtion coefficients $\alpha_{\lambda}$ in continuum spectra often have the vast regions of smooth dependence
on the wavelength \citep[see] []{gra76}. In these regions they can be approximated by power law
$\alpha_{\lambda}\simeq \alpha(T)\lambda^n $. Then the values $b_{\lambda}(T_0)$ and $a_{\lambda}(T_0)$
(see Eq.(10)) are independent of $\lambda $: $b_{\lambda}(T_0)=b(T_0)$ and $a_{\lambda}(T_0)=a(T_0)$.
In some cases the absorbtion coefficient has the form of product $\alpha_0(\lambda)$ and $\alpha_1(T)$, i.e.
$\alpha_{\lambda}(T)=\alpha_0(\lambda)\alpha_1(T)$. Very important example of such dependence
is the case of bound-free transitions in negative hydrogen ion $H^{-}$
(see in more detail below). In such cases the values $a_{\lambda}(T)$ and $b_{\lambda}(T)$ also are independent of
wavelength.

Using the relations (17), Eq.(13) can be written in the form:

\[
 \langle H_{\lambda}(\eta)\rangle =H^{(0)}_{\lambda}+
\]
\[
2\pi\eta^2\int^{\infty}_0d\tau_{\lambda}\,E_2(\tau_{\lambda})
\left(10B_{\lambda}+5\lambda\,\frac{\partial
 B_{\lambda}}{\partial\lambda}
+\frac{\lambda^2}{2}\,\frac{\partial^2 B_{\lambda}}{\partial\lambda^2}\right)+
\]
\begin{equation}
2\pi\eta^2\int^{\infty}_0d\tau_{\lambda}
\,E_2(\tau_{\lambda})b(T_0)\left(5B_{\lambda}(T_0)
+\lambda \,\frac{\partial B_{\lambda}}
 {\partial\lambda}\right)+ A_{\lambda}.
\label{18}
\end{equation}

The second term in the right part of this expression describes the contribution of temperature fluctuation
to the spectrum due to influence of Planck's function, i.e. this is the contribution of value
$\langle B_{\lambda}\rangle$. The estimate of this contribution from model spectra, for example, from
models of \citep{kur74}, is of interest for knowledge of influence of temperature fluctuations on
continuum spectra. Beforehand, note that the contribution of $\langle B_{\lambda}\rangle$ is smaller than that
of term with $b(T_0)$.

Because $\tau_{\lambda}$ and $E_2(\tau_{\lambda})$ depend on wavelength $\lambda$, the $\lambda$ -
differentiation cannot be taken out the integral in Eq.(18). But for absorbtion coefficients, mentioned above,
exist the relations:

\[
\frac{\partial \alpha_{\lambda}}{\partial \lambda} =\beta(\lambda)\alpha_{\lambda},\,\,\,
\frac{\partial \tau_{\lambda}}{\partial \lambda}=\beta(\lambda)\tau_{\lambda},
\]
\[
\beta(\lambda) =\frac{1}{\alpha_0(\lambda)}
\frac{\partial \alpha_0(\lambda)}{\partial \lambda},
\]
\noindent which lead to the following formulas:

\begin{equation}
2\pi\lambda\,\int^{\infty}_0d\tau_{\lambda}\,E_2(\tau_{\lambda})\frac{\partial B_{\lambda}}{\partial \lambda}
=\lambda\frac{\partial H^{(0)}_{\lambda}}{\partial\lambda}-\lambda\beta(\lambda)T_{\lambda},
\label{19}
\end{equation}

\begin{equation}
T_{\lambda}=2H^{(0)}_{\lambda}-2\pi I^{(0)}_{\lambda}(0,1)=4\pi\int^1_0d\mu\,
\mu[I^{(0)}_{\lambda}(0,\mu)-I^{(0)}_{\lambda}(0,1)],
\label{20}
\end{equation}

\[
2\pi\lambda^2\int^{\infty}_0d\tau_{\lambda}E_2(\tau_{\lambda})\frac{\partial^2 B_{\lambda}}{\partial\lambda^2}=
\lambda^2\frac{\partial^2 H^{(0)}_{\lambda}}{\partial\lambda^2}-\lambda^2\,
\frac{\partial\beta(\lambda)}{\partial \lambda}\,T_{\lambda}
\]
\begin{equation}
-2\lambda^2\beta(\lambda)\frac{\partial T_{\lambda}}{\partial\lambda}+\lambda^2\beta^2(\lambda)M_{\lambda}.
\label{21}
\end{equation}
\noindent When deriving these formulas, we used Eqs.(1) and (3). The values $\beta(\lambda)$ and
$\partial \beta/\partial \lambda$ for negative hydrogen ion are presented in Table 1. They were calculated
numerically from figure for $\alpha_{\lambda}$, given in \citep{gra76}.

Function $T_{\lambda}$ depends on the flux of outgoing radiation $H^{(0)}_{\lambda}$ and the intensity of radiation
$I^{(0)}_{\lambda}(0,1)$ escaping perpendicular to the surface. Note that due to the law of radiation darkening
($I^{(0)}_{\lambda}(0,\mu)<I^{(0)}_{\lambda}(0,1)$), the function $T_{\lambda}$ is negative.
This function can be calculated from a model of photosphere. Here we will use the simplest model, assuming
that the function $B_{\lambda}(\tau_{\lambda})$ increases linearly with $\tau_{\lambda}$.  This model is presented in
a number of books \citep[see, for example,] []{sob69}.

Below we restrict ourselves by the Wien limit of a spectrum, when the parameter $g_0/\lambda T >>1$. In this case
simple model, mentioned above, gives rise to the relations:

\[
T_{\lambda}\simeq - H^{(0)}_{\lambda},
\]

\begin{equation}
M_{\lambda}=2T_{\lambda}+2\pi\,\int^{\infty}_0d\tau_{\lambda}B_{\lambda}(T_0)(\tau_{\lambda}-1)
\exp{(-\tau_{\lambda})}\simeq H^{(0)}_{\lambda}.
\label{22}
\end{equation}

According to formulas (19) and (21), contribution of $\langle B_{\lambda}(\tau_{\lambda})\rangle$
into $\langle H_{\lambda}(\eta)\rangle$ acquires the form:

\[
\langle H_{\lambda}(\eta)\rangle =H^{(0)}_{\lambda}+\eta^2\left[\left(10H^{(0)}_{\lambda}+5\lambda\,
\frac{\partial H^{(0)}_{\lambda}}{\partial \lambda}+\frac{1}{2}\lambda^2\,
\frac{\partial^2 H^{(0)}_{\lambda}}{\partial \lambda^2}\right)-\right.
\]
\begin{equation}
\left.\lambda^2\beta(\lambda)\,\frac{\partial T_{\lambda}}{\partial \lambda}
-\left(5\lambda \beta(\lambda)+\frac{1}{2}
\lambda^2\frac{\partial\beta(\lambda)}{\partial \lambda}\right)T_{\lambda}
+\frac{1}{2}\lambda^2\beta^2(\lambda)M_{\lambda}
\right].
\label{23}
\end{equation}
\noindent This formula depends on the specific form of absorbtion coefficient. Because of the difference
$\langle B_{\lambda}(T_0,\eta)\rangle -B_{\lambda}(T_0)$ is positive and slowly decreases with grow of wavelength,
then the term $\langle B_{\lambda}\rangle$ at $\eta^2$ gives the positive contribution to observed spectrum
 $\langle H_{\lambda}(\eta)\rangle$. Yet the contribution of term $b_{\lambda}(T_0)$, in principle, can exceed
the contribution from $\langle B_{\lambda}(T_0)\rangle$, and, in particular, gives rise to negative correction to
 $\langle H_{\lambda}\rangle$.

\subsection{Estimate of expressions with $a_{\lambda}(T_0)$}

The term $A_{\lambda}$ in formula (14) describes the change of spectrum  as a result of difference between
$\langle\alpha_{\lambda}\rangle$ and $\alpha_{\lambda}(T_0)$ (see formula (11)).
Negative part of $A_{\lambda}$ presents the decrease of flux due to additional term with
$\eta^2$ in $\exp(-\langle\tau_{\lambda}\rangle)$ (see (12)). The positive part describes the increasing of flux
due to effective source $\langle\alpha_{\lambda}B_{\lambda}\rangle$. The difference of these terms is lower than
every separate term.

It is of interest to estimate the contribution of term $A_{\lambda}$ to expressions (13) and (18). Let us replace
the terms $a_{\lambda}(T_0)$ and $a_{\lambda}(T'_0)$ by effective mean value $\overline{a}_{\lambda}$,
taken at the level $\tau_{\lambda}\simeq 2/3$, where is located the main source of radiation. As a result, one
can obtain the expression:

\[
 A_{\lambda}\simeq 2\pi\eta^2\overline{a}_{\lambda}\int^{\infty}_0d\tau_
{\lambda}\left(2E_2(\tau_{\lambda})
 -e^{-\tau_{\lambda}}\right)B_{\lambda}(T_0)=
\]
\[
-2\pi\eta^2\overline{a}_{\lambda}\int^{\infty}_0d\tau
 {\lambda}\tau_{\lambda}E_2(\tau_{\lambda})
\frac{\partial B_{\lambda}(T_0)}{\partial \tau_{\lambda}}
\equiv
\]
\begin{equation}
\eta^2\overline{a}_{\lambda}T_{\lambda}
\simeq-\eta^2\overline{a}_{\lambda}\delta_{\lambda}H^{(0)}_{\lambda}
\label{24}
\end{equation}
\noindent This expression is less than the term  $10H^{(0)}_{\lambda}$ in Eq. (23) if the value
$\overline {a}_{\lambda}$ is less or of the order of unity.  In these cases the term $A_{\lambda}$ can be neglected.
The coefficient $\delta_{\lambda}$, according to approximate Sobolev's (1969) formula, is estimated as the ratio of
the mean absorbtion factor over all spectrum to the local coefficient $\alpha_{\lambda}$. In the case of absorbtion
by negative hydrogen ions $H^{-}$ the Chandrasekhar's estimation  \citep[see] []{cha50} gives the value
 $\delta_{\lambda}\simeq 0.26$.
This gives the estimation $A_{\lambda}\simeq (1\div 2)H^{(0)}_{\lambda}$, which is less than the term
$10 H^{(0)}_{\lambda}$ in formula (23).

\section{Estimates in Wien's limit}

The case $g_0/(\lambda T_0)=14388/(\lambda T_0)\gg 1$ corresponds to Wien's limit of thermal emission.
Recall that at the maximum of Planck's emission $\lambda_{max}T=2900$. This gives $g_0/(\lambda_{max}T)\simeq 5$, i.e.
the Wien approximation  $B_{\lambda}(T)=(2hc^2/\lambda^5)\exp{(-g_0/\lambda T)}$ exists at least up to
maximum of spectrum. The relative deviation from Planck's formula at maximum of emission consists of $0.7\,\%$.
At $g_0/\lambda T=3.6$ this deviation is  $\approx 2.8\,\%$, that corresponds to the value
 $\lambda(\mu m) T\simeq 4000$.

This means that the Wien limit occurs practically up to $\lambda =0.5$ $\mu m$ at $T=8000$ K, and up
 to $\lambda\simeq 1$
$\mu m$ at $T\simeq 8000$ K. Note that in this temperature interval the absorbtion coefficient in continuum, basically,
is due to bound-free transitions of outer electron in negative hydrogen ion $H^{-}$ \citep[see] []{cha50}.
Below we will give the formulas of this case, in detail.
In Wien's limit the Einshtein-Milne correction due to stimulated emission (the factor $(1-\exp{(-g_0/\lambda T)})$ is
low and the formulas for absorbtion coefficient \citep[see] []{gra76} become simpler.

For the absorbtion factors of the type

\begin{equation}
\alpha_{\lambda}=\alpha_0(\lambda)T^{\gamma}e^{-g_1/T}
\label{25}
\end{equation}
\noindent dimensionless coefficients $b(T)$ and $a(T)$, according to Eq. (10), acquire the form:

\[
 b(T)=\gamma+\frac{g_1}{T},
\]
\begin{equation}
 a(T)=\frac{\gamma(\gamma-1)}{2} +\frac{(\gamma-1)g_1}{T}+\frac{g^2_1}{2T^2}.
\label{26}
\end{equation}

Formula (25) is valid for the description bound-free transitions in  $H^{-}$, where $\gamma=-1.5$ and
$g_1=-8700$ (if the wavelengths are in microns).

The coefficients $b(T)$ and $a(T)$ obey power-law dependence on the temperature.  Therefore in Wien's approximation
Eq.(18) (without the term $A_{\lambda}$) can be simplified  if we transform the terms of type $B_{\lambda}(T_0)/T^n_0$
to expressions with the differentiation over wavelength. For this aim we can use the following equalities
(introducing the notion $f_1(\lambda T)=\exp(-g_1/\lambda T)$):

\begin{equation}
 \frac{f_1(\lambda T)}{T}=\frac{\lambda^2}{g_1}\,\frac{\partial f_1}{\partial\lambda},\,\,\,\,\,\,
 \frac{f_1(\lambda T)}{T^2}=\left(\frac{\lambda^2}{g_1}\right)^2\left[\frac{2}{\lambda}\,\frac{\partial f_1}
 {\partial\lambda}
 +\frac{\partial^2 f_1}{\partial\lambda^2}\right].
\label{27}
\end{equation}
\noindent Taking into account, that in Wien's approximation $B_{\lambda}\sim \lambda^{-5}\exp(-g_0/\lambda T)$,
we derive the relations:

\[
 T\frac{\partial B_{\lambda}}{\partial T}=\frac{g_0}{\lambda T}B_{\lambda},\,\,\,\, \frac{B_{\lambda}}{T}=
 \frac{\lambda}{g_0}\left(5B_{\lambda}+\lambda\,\frac{\partial B_{\lambda}}{\partial\lambda}\right),
\]
\begin{equation}
 \frac{B_{\lambda}}{T^2}=\frac{\lambda^2}{g^2_0}\left[30B_{\lambda}+12\lambda\,\frac{\partial B_{\lambda}}{\partial\lambda}
 +\lambda^2\,\frac{\partial^2 B_{\lambda}}{\partial\lambda^2}\right].
\label{28}
\end{equation}

Using these relations, and Eqs. (19)  and (21),  we derive Eq. (18)  (without the term  proportional to $a(T)$)
in Wien' approximation:

\[
 \langle H_{\lambda}(\eta)\rangle=H^{(0)}_{\lambda}+
\]
\[
\eta^2\left[C_0(\lambda)H^{(0)}_{\lambda}
+ D_0(\lambda)T_{\lambda}+G_0(\lambda)M_{\lambda}
 +C_1(\lambda)\lambda\frac{\partial H^{(0)}_{\lambda}}{\partial \lambda}+\right.
\]
\begin{equation}
\left.D_1(\lambda)\,\lambda\,\frac{\partial T_{\lambda}}{\partial \lambda}+
 C_2(\lambda)\,\lambda^2\,\frac{\partial^2 H^{(0)}_{\lambda }}{\partial \lambda^2}\right].
\label{29}
\end{equation}

\[
C_0(\lambda)=10+5\gamma +30\xi,\,\,\,\,\,\,C_1(\lambda)=5+\gamma+12\xi,
\]
\[
\,\,\,\,\,\,\,\,\,\,\,C_2(\lambda)=(0.5+\xi),
\]

\[
D_0(\lambda)=-\lambda\beta(\lambda)C_1-\lambda^2\frac{\partial\beta(\lambda)}{\partial \lambda}\,C_2,
\]
\begin{equation}
G_0(\lambda)=\lambda^2\beta^2(\lambda)C_2,\,\,\,\,
D_1(\lambda)=-2\lambda\beta(\lambda)C_2.
\label{30}
\end{equation}
\noindent Here we use the dimensionless values $\xi=g_1\lambda /g_0$ with $g_0=14388 $. For the case
 $\alpha_{\lambda}(H^{-})$ one has $\gamma=-3/2,\, g_1=-8700,\,g_1/g_0=-0.6$ and $\xi=-0.6\lambda $.
As it was mentioned earlier, we can neglect the term $A_{\lambda}$ if the absorbtion in negative hydrogen ions
are considered.
 According to \citep{cha50, sob69}  this case is valid for the solar type stars with
$T_e<7\cdot 10^3$ K.

In Table 2 we present the corrections to $\langle H_{\lambda}(\eta)\rangle$
for a number of solar type stars ($\beta$ Hyi (G2IV), $\mu$ Vel (G5III), $\iota$ Per (G0V))
and the Sun, using the catalog \citep{ale96}. The corrections were calculated for
 $\lambda =0.55\mu$m and 0.6 $\mu$m. For these wavelengths one can neglect the contribution of free-free transitions
in $H^{-}$. Note that the columns with positive numbers in  Table 2  refer to $\Phi^{(0)}_{\lambda}$, and the columns
with negative numbers present the values $\Phi_{\lambda}$.  The notions of these values are given below.

Formula (29), with taking into account Eq.(22), acquires the form:

\[
 \langle H_{\lambda}(\eta)\rangle\equiv H^{(0)}_{\lambda}(1
+\eta^2\Phi_{\lambda})=H^{(0)}_{\lambda}+
\eta^2\left\{\Phi^{(0)}H^{(0)}_{\lambda}+\right.
\]
\begin{equation}
 \left.A_0 H^{(0)}_{\lambda}+A_1 \lambda
\frac{\partial H^{(0)}_{\lambda}}{\partial\lambda}+\xi\lambda^2
 \frac{\partial^2 H^{(0)}_{\lambda}}{\partial\lambda^2}\right\},
\label{31}
\end{equation}
\noindent where we introduce the notion $\beta'=\partial\beta/\partial\lambda$, and the factors $A_0$ and $A_1$
 have the following form:

\[
A_0=(5+\lambda\beta)\gamma+\xi(30+12\lambda\beta+\lambda^2(\beta^2+\beta')),
\]

\begin{equation}
A_1=\gamma+\xi(12+2\lambda\beta).
\label{32}
\end{equation}

\noindent The term $\Phi^{(0)}H^{(0)}_{\lambda}$ coincides with that in square brackets in Eq. (23), if the
relations (22) are used. This term has the form:

\[
\Phi^{(0)}_{\lambda}H^{(0)}_{\lambda}=\left[[10+5\lambda\beta+\frac{\lambda^2}{2}(\beta^2+\beta')]H^{(0)}_{\lambda}
\right.
\]
\begin{equation}
\left.+(5+\lambda\beta)\lambda\frac{\partial H^{(0)}_{\lambda}}{\partial\lambda}
+\frac{\lambda^2}{2}\frac{\partial^2 H^{(0)}_{\lambda}}{\partial\lambda^2}\right].
\label{33}
\end{equation}
\noindent  Recall that this term presents  contribution of $\langle B_{\lambda}\rangle$.

\subsection{Results of calculations and the discussions}

Usually the star fluxes are presented in logarithmic scale $\lg{H_{\lambda}}$ or in star magnitudes
$m_{\lambda}=-2.5\lg{H_{\lambda}}$. Our formulas have the derivatives
$\partial H_{\lambda}/\partial \lambda$ and $\partial^2 H_{\lambda}/\partial \lambda^2$. Differentiation of the
relation $\ln{H_{\lambda}}=\ln{10}\,\lg{H_{\lambda}}$ gives rise to expressions:

\begin{equation}
\frac{\partial H_{\lambda}}{\partial \lambda}=H_{\lambda}\ln{10}\,\frac{\partial \lg{H_{\lambda}}}{\partial \lambda}=
-H_{\lambda}\,\frac{\ln{10}}{2.5}\,\frac{\partial m_{\lambda}}{\partial \lambda},
\label{34}
\end{equation}

\[
\frac{\partial^2 H_{\lambda}}{\partial \lambda^2}=H_{\lambda}\left[\ln{10}\,\frac{\partial^2 \lg{H_{\lambda}}}
{\partial \lambda^2}+\left(\ln{10}\,\frac{\partial \lg{H_{\lambda}}}{\partial \lambda}\right)^2\,\right]=
\]
\begin{equation}
H_{\lambda}\left[\left(\frac{\ln{10}}{2.5}\,\frac{\partial m_{\lambda}}{\partial \lambda}\right)^2 -
\frac{\ln{10}}{2.5}\,\frac{\partial^2 m_{\lambda}}{\partial \lambda^2}\right].
\label{35}
\end{equation}

These formulas show that the derivatives are proportional to $H_{\lambda}$. It means that the expression (31)
may be presented in the form:

\begin{equation}
\frac{\langle H_{\lambda}(\eta)\rangle - H^{(0)}_{\lambda}}{H^{(0)}_{\lambda}}\equiv
 \frac{\Delta H_{\lambda}}{H^{(0)}_{\lambda}}
=\eta^2 \Phi_{\lambda},
\label{36}
\end{equation}
\noindent
The value $\Delta H_{\lambda}$ can be presented as a corresponding change $\Delta T$ \citep[see][]{sob69}.
As a result, we derive the estimate:

\begin{equation}
\Delta T\simeq (2.8\div 4)\left(\frac{g_0}{\lambda T_0}\right)^{-1}\eta^2 \Phi_{\lambda}T_0,
\label{37}
\end{equation}
\noindent  Instead of  $T_0$ in such estimates one may be taken the effective temperature $T_e$ in the regions, where
 $B_{\lambda}(T_e)$ satisfactory describes the spectrum. Recall that in Wien's region even low temperature changes
 $\Delta T$ give rise to fairly high changes of fluxes.

Let us present the explicit formula for function $\Phi_{\lambda}$, if the absorbtion coefficient has form (25).
This formula follows directly from Eqs. (31) - (35):

\[
\Phi_{\lambda}=10+5\lambda\beta+\frac{\lambda^2}{2}(\beta^2+\beta')+A_0+
\]
\[
(5+\lambda\beta +A_1)\lambda\ln{10}\,\frac{\partial \lg{H_{\lambda}}}{\partial \lambda}+
\
\]
\begin{equation}
(\xi+0.5)\,\lambda^2\,\left[\ln{10}\,\frac{\partial^2 \lg{H_{\lambda}}}
{\partial \lambda^2}+\left(\ln{10}\,\frac{\partial \lg{H_{\lambda}}}{\partial \lambda}\right)^2\right].
\label{38}
\end{equation}

First of all, we obtain the part of expression $\Phi_{\lambda}$, which is due to contribution of mean Planck's
function $\langle B_{\lambda}\rangle$. This expression can be obtained from Eq. (38), where terms with
 $\gamma$ and $\xi$ have to be omitted. Let us denote this expression as $\Phi^{(0)}_{\lambda}$. To calculate
this term one can use the model spectra, or use the observed spectra $\langle H_{\lambda}(\eta)\rangle$, assuming
that at low values $\eta$ exists approximate equality $\langle H_{\lambda}(\eta)\rangle\simeq H^{(0)}_{\lambda}$.

As it was mentioned earlier, the  spectra depend strongly on the contribution of the coefficient
 $b_{\lambda}(T_0)$ (see Eq. (10)), which describe the change of effective source
$\langle \alpha_{\lambda}B_{\lambda}\rangle$ due to fluctuations of first temperature derivatives in absorbtion factor
$\alpha_{\lambda}(T)$ and in Planck's function $B_{\lambda}(T)$. As a role, the first derivatives  characterize
the changes most sensitively than the second ones. The contribution of second order
derivatives give rise to less contribution to changes in spectra. For this reason, the calculation of
 $\Phi^{(0)}_{\lambda}$ is useful for comparison  with basic terms proportional to
mean value of the product of first order derivatives.

It is of interest, that calculations, presented in Table 2, show that the terms with  $\beta$ and $\beta^{'}$
 give comparatively low contribution  to  $\Phi_{\lambda}$ , less than $(2\div 5)\%$ compared with other terms.
Note, that this low contribution is not the consequence of smallness of $\beta$ and $\beta^{'}$ (see Table 1 )
,  and basically are due to powers of small wavelengths
($\lambda\sim 0.5$ $\mu m$) at coefficients $\beta $ and $\beta^{'}$.
Therefore for estimations of $\langle B_{\lambda}\rangle$ (term $\Phi^{(0)}_{\lambda}$) can be used the simple
formula:

\begin{equation}
\Phi^{(0)}_{\lambda}H^{(0)}_{\lambda}\simeq 10H^{(0)}_{\lambda}
+5\lambda \frac{\partial H^{(0)}_{\lambda}}{\partial \lambda}
+\frac{\lambda^2}{2}\frac{\partial^2H^{(0)}_{\lambda}}{\partial\lambda^2}.
\label{39}
\end{equation}
\noindent  The term with  $b_{\lambda}$ can be simplified accordingly. The sense of these  simplifications
consists of conclusion, that to obtain the estimates in Eq. (18), one can take out of the integrals the
differentiations over wavelength. Apparently, this can be used in other cases, if the absorbtion coefficients
change slowly in the range of considering spectrum, and if the spectrum itself has fairly smooth form.
In this case one can introduce the large-scale differentiation over wavelengths for spectrum itself, and use
the averaged absorbtion factor.

In Table 3 we presented the values of functions $\Phi^{(0)}_{\lambda}$ and $\Phi_{\lambda}$
for stars $\alpha$ PsA and $\beta$ Ari, calculated from homogeneous catalog \citep{ale96}.
The tables of star magnitudes $m_{\lambda}$ in this catalog are presented with the interval
$\Delta \lambda =0.0025$ $\mu$m. The first star is of the class A3V with $T_e\simeq 8300$ K, and the second one is
of the class A5V with $T_e\simeq 8500$ K.
Spectra of these stars does not determined by absorbtion in negative ion $H^{-}$. To receive the esimates
of temperature fluctuations influences we used the absorbtion coefficients in continuum given in \citep{all73}.
We used the parabolic interpolation of data from this book for intervals  $\lambda =0.4, 0.5, 0.8$ $\mu$m
and for $\Theta =5040/T=0.4, 0.5, 0.6$ at the value of electron pressure  $\lg P_e=1$.
As a result, we derived the formula:

\[
\alpha_{\lambda}(\Theta)= 10^{C(\lambda, \Theta)},
\]
\[
C(\lambda, \Theta)=\frac{25}{3}[\lambda^2(20\Theta^2-21\Theta+5.3)+
\]
\begin{equation}
\lambda(-24\Theta^2+25.5\Theta-6.33)
+(5.8\Theta^2-6.18\Theta+1.468)].
\label{40}
\end{equation}
\noindent Taking into account that $\lambda$ - dependence in this expression  is fairly smooth, we used for
estimations the mentioned above (see Eq. (39) and so on) simple expressions.

The negative corrections $\Phi_{\lambda}$ denote that the temperature (see the notions (36) and (37))
is less compared with the temperature obtained from photosphere models,
which do not take into account the temperature fluctuations. Especially important conclusion from our estimates
is that most considerable corrections to  $\langle H_{\lambda}(\eta)\rangle$ are due to terms with
$b_{\lambda}(T)$ (see Eq. (10)). These terms give rise to negative corrections whereas the contribution of
$\langle B_{\lambda}(T)\rangle$ gives positive corrections. The contribution of terms with $b_{\lambda}$
 is determined by the  average of product of first temperature derivative from absorbtion factor and from the Planck
function, i.e. most important linear expansions over $T'$ are averaged.

It is of interest to note that observed Sun's spectrum in ultraviolet region is lower than that determined by
accepted effective temperature \citep{sti91, chn10}. Usually this effect is related with
influence of absorbtion of metals. However, the possible contribution of temperature fluctuations is also not
eliminated. Though this mechanism is not studied in detail, but preliminary we can note that the calculations
of source function of new transfer equation (4) in ultraviolet region demonstrate the strong difference from the usual
Planck's function.

The estimates (37) for the Sun at $\lambda =0.55$ $\mu$m and $T_e\simeq 5770$ K give
$\Delta T\simeq -(2.8\div 4)\eta^2\cdot 9.8 T_e\simeq -(2.8\div 4)\cdot 12470\eta^2$, i.e.
$\Delta T\simeq -(31\div 45)$ K
for $\eta =0.03$, and  $\Delta T\simeq -(87\div 125)$ K for $\eta=0.05$. In percents these estimates are
 $(0.5\div 0.8)\%$ and $(1.5\div 2.1)\%$ of $T_e$, correspondingly.

These estimates are given for $\lambda\sim 0.55$ $\mu$m and practically do not depend on low changes in taking
$T_0\sim T_e$. Note, that to estimate the change in effective temperature of a star, one has calculate the
change of all the photosphere spectrum in the range of all wavelengths.

\section{ Influence of temperature fluctuations on color indices of radiating objects}

The color index of observing objects is characterized by the difference of radiation fluxes between standard
wavelength intervals.
Usually the observed fluxes  $F_{\lambda}$ correspond to standard wavelength intervals near the central
wavelengths ($\lambda_U=0.36\mu$m, $\lambda_B=0.44\mu$m, $\lambda_V=0.55\mu$m, $\lambda_R=0.7\mu$m).
For example, the color index $U-B$ is determined according to formula:

\begin{equation}
U^{(0)}-B^{(0)}=-2.5\lg\frac{F^{(0)}_U}{F^{(0)}_B}+ Const.
\label{41}
\end{equation}
\noindent The value  $Const$ is taken from the reasons of convenience. Frequently one takes $Const=0$ for
stars of A0V-type.

Usually one considers that fluxes $F^{(0)}_{\lambda}=B_{\lambda}(T_0)$, where temperature $T_0$ on an average
describes the observed part of spectrum. This relation formally follows from usual transfer equation at
supposition that the source of radiation $B_{\lambda}(T_0)$ is independent of  optical depth in radiating
region. The radiating objects can have the fluctuating temperature. In this case the source is described by
the formula
$\langle \alpha_{\lambda}(T)B_{\lambda}(T)\rangle/\langle B_{\lambda}(T)\rangle \equiv
S_{\lambda}(T_0,\eta)/\langle B_{\lambda}(T)\rangle$
(see Eqs. (4) and (5)).

As we know, in this case the flux $\langle H_{\lambda}(\eta)\rangle\equiv \langle F_{\lambda}\rangle$
depends on the degree of temperature fluctuations  $\eta$. Assuming that this effective source is independent of
optical depth, we obtain the expression
$\langle F_{\lambda}\rangle=S_{\lambda}(T_0,\eta)/\langle B_{\lambda}(T)\rangle$.
This means that the color indices $\langle U\rangle-\langle B\rangle$ and so on depend on degree of fluctuations
$\eta$. As usually, we restrict ourselves to low fluctuations $\eta^2\ll 1$ and instead of Eq. (41) obtain the relation:

\begin{equation}
\langle U\rangle -\langle B\rangle =U^{(0)}-B^{(0)}+ \eta^2 \Delta(U-B),
\label{42}
\end{equation}
\noindent where $\Delta(U-B)$ is expressed by the formula:

\[
\Delta(U-B) =-2.5(K_{\lambda_U}-K_{\lambda_B}),
\]
\[
K_{\lambda}=\frac{T_0^2}{2B_{\lambda}(T_0)}\cdot \frac{\partial^2 B_{\lambda}(T_0)}{\partial T^2_0}+
\frac{T_0}{B_{\lambda}(T_0)}\frac{\partial B_{\lambda}(T_0)}{\partial T_0}\cdot b_{\lambda}(T_0)
\]
\begin{equation}
\equiv K^{(0)}_{\lambda}+K^{(1)}_{\lambda}\,b_{\lambda}(T_0).
\label{43}
\end{equation}
\noindent Analogous formulas characterize the color indices $B-V$ and $V-R$. The numerical values for coefficients
$K^{(0)}_{\lambda}(T_0)$ and $K^{(1)}_{\lambda}(T_0)$ are presented in Table 4 for wavelengths
$\lambda=0.36; 0.44; 0.55$ and $0.7$ $\mu$m. Note, that these factors are positive. At growing of wavelengths
these coefficients  diminish.

It is seen from Eq.(43) that the color indices strongly depend on specific form of absorbtion coefficients.
This is seen clearly from the example, when we use the absorbtion factor (25) for bound - free transitions in
negative hydrogen ions. In this case $b_{\lambda}(T_0)$ has the form (26) with $\gamma=-1.5$ and
$g_1=-8700$. The term $K^{(0)}_{\lambda}$ describes the influence of temperature fluctuations without contribution
of absorbtion factor. This term is positive, as it was mentioned earlier. The expressions $\Delta(U-B), \Delta(B-V)$,
and  $\Delta(V-R)$ with allowance for only $K^{(0)}_{\lambda}$ - term are negative for all mean temperatures $T_0$.
The consideration of absorbtion coefficients gives rise to the result that the total coefficient
$K^{(0)}_{\lambda}+K^{(1)}_{\lambda}\,b_{\lambda}(T_0)$ strongly differs from the term $K^{(0)}_{\lambda}$
(see Table 5). In Table 5 we also present the values of $\Delta(U-B)$ and so on, which take into account
 both coefficients -
$K^{(0)}_{\lambda}$ and $K^{(1)}_{\lambda}$.

At comparatively low temperatures the correction to color indices is negative. Then, with the grow of the temperature,
this correction acquires positive sign. Most high the influence of temperature fluctuation is in color index
 $U-B$, i.e. for small wavelengths. It seems, that in active galactic nuclei the temperature fluctuations
may be high. In these cases the consideration  of fluctuations can change the color indices considerably.
Besides, at introduction of color indices would be useful, though roughly, take into account the temperature
gradient near the surface of emitting object, for example, using the mentioned approximate formula in \citep{sob69}.
It appears the inverse problem of estimation of degree of temperature fluctuations $\eta$ is simpler from
consideration of color indices of cosmic objects with like physical conditions.

\section{Conclusion}

Let us present short discussion of obtained results. The existence of stochastic processes in photospheres of stars,
active galactic nuclei etc., which arises
due to convective and turbulent motions, generation and evolution of magnetic fields, give rise to
stochastic behavior of temperature, intensity of radiation and  absorbtion coefficients.
These stochastic
processes determine the mean radiation flux $\langle H_{\lambda}\rangle$. This flux differs from the flux
 $H^{(0)}_{\lambda}$, which takes into account only dependence of all values from mean temperature $T_0$
and is independent of temperature fluctuations $T'$. The numerous models of photospheres determine the mean
temperature as a function of distance from the photosphere's surface.

Assuming the degree of temperature fluctuations $\eta =\sqrt{\langle T'^2\rangle}/T_0$ as a low value, we obtain
the relation between
the observed flux $\langle H_{\lambda}\rangle $ and model flux $H^{(0)}_{\lambda}$. The difference between these
fluxes is proportional to $\eta^2$ and can assume both positive and negative values.
This is determined by the specific form of absorbtion coefficient. In Wien's limit this difference can be estimated
from the observed spectrum, because it depends of the first and the second derivatives over wavelength, which can be
calculated numerically directly from the observed spectrum and is independent of specific choice of photosphere
model.

We calculated this difference of fluxes for the Sun, and solar type stars ($\beta$ Hyi, $\mu$ Vela and $\iota$ Per)
for $\lambda=0.55\mu$m and $\lambda=0.6\mu$m, using the homogeneous Pulkovo spectrophotometric catalog
 \citep[see] []{ale96}.
Also we obtain the corresponding estimates for stars $\alpha$ PsA and $\beta$ Ari.

These estimates demonstrate that the most high correction is due to mean value of product of first temperature
derivatives of the absorbtion coefficient and the Planck function.  The estimations show that the difference
of spectra $\langle H_{\lambda}\rangle$ and  $H^{(0)}_{\lambda}$ is fairly high and can be used for
correction of model results, especially in ultraviolet region of spectra.
We also described the influence of temperature fluctuations on the color indices. It appears these corrections
can be high, especially for active galactic nuclei.

{\bf Acknowledgements.}
This research was supported by the program of Prezidium of RAS
��- 21, the program of the Department of Physical Sciences of RAS
��- 16, by the Federal Target program " Scientific and scientific-
pedagogical personnel of innovative Russia"  2009 -2013 and the Grant
from President of the Russian Federation "The basic Scientific Schools"
NSh-1625.2012.2.

Authors are very grateful to anonymous referee for very useful remarks.

\clearpage

\begin{table}
\caption[]{\small  Functions $\beta(\lambda)$ and $\partial\beta(\lambda)/\partial \lambda$ for absorbtion
 factor $\alpha_{\lambda}(H^{-})$.}
\label{tab1}
\vspace{6mm}
 \centering
\begin{tabular}{llllllllllll}
 \hline
 $\lambda$,\,\,$\mu $m             & 0.2 & 0.25 & 0.3 & 0.35 & 0.4 & 0.45  & 0.5  & 0.55 & 0.6 & 0.65 & 0.7 \\
 \hline
 $\beta$                           & 3.5 & 2.4 & 1.7 & 1.28 &  1   & 0.76  & 0.58 & 0.41 & 0.3  & 0.22 & 0.15 \\
 $\partial\beta/\partial\lambda$   & -28 & -18 & -11 & -6.8 & -5.3 & -4.25 & -3.5 & -2.7 & -2.0 & -1.5 & -1.25 \\
\hline
\end{tabular}
\end{table}

\begin{table}
\caption[]{\small  Functions $\Phi^{(0)}_{\lambda}$ and $\Phi_{\lambda}$ for the solar type stars.}
\label{tab2}
\vspace{6mm}
 \centering
\begin{tabular}{lllllllll}
\hline
                                  &  the Sun  & &$\,\,\beta$ Hyi  & & $\,\,\mu$ Vel & & $\iota$ Per&  \\

\hline
 $\lambda=0.55\mu$m                 & 4.3 & -9.8  & 3.6 & -23.6 & 7.9 & -12.9 & 5.9 & -18.0 \\
 $\lambda=0.6\mu$m                 & 2.7 & -10.2 & 3.3 & -26.4 & 3.9 & -14.4 & 5.6 & -20.1 \\

\hline
\end{tabular}
\end{table}

\begin{table}
\caption[]{\small  Functions $\Phi^{(0)}_{\lambda}$ and $\Phi_{\lambda}$ for spectra
 of $\alpha $ PsA and $\beta $ Ari.}
\label{tab3}
\vspace{6mm}
 \centering
\begin{tabular}{llllll}
\hline
 $\lambda$,\,\,$\mu $m                     & 0.45 & 0.47 & 0.50 & 0.55 & 0.60  \\
 \hline
 $\alpha$ PsA,\,$\Phi^{(0)}_{\lambda}$       &  1.7  & 1.4   &  1.0   & 0.4   & 0.0  \\
 $\alpha$ PsA,\,$\Phi_{\lambda}$           & -10.2 & -11.3 & -12.2  & -13.5 & -13.4  \\
 \hline
 $\beta$ Ari, \,$\Phi^{(0)}_{\lambda}$     &  3.1  &  2.8  &  1.9   &  1.4  & 0.9 \\
 $\beta$ Ari, \,$\Phi_{\lambda}$           & -8.6  & -9.2 &  -10.8  & -11.0 & -11.1 \\
\hline
\end{tabular}

\end{table}

\begin{table}
\caption[]{\small The values of $K^{(0)}_{\lambda}(T_0)$ and $K^{(1)}_{\lambda}(T_0)$.}
\label{tab4}
\vspace{6mm}
\centering
\begin{tabular}{llllllllllll}
\hline
 $T_0$, \,$K$ &4$\cdot10^3$ &5$\cdot10^3$ &6$\cdot10^3$ &7$\cdot10^3$ &8$\cdot10^3$ &9$\cdot10^3$ &10$\cdot10^3$
  &12$\cdot10^3$ &14$\cdot10^3$ &16$\cdot10^3$ &18$\cdot10^3$\\
\hline
 $K^{(0)}_{\lambda_U}$   & 39.93 & 23.98 & 15.60 & 10.73 & 7.71 & 5.72 & 4.37 & 2.73 & 1.82 & 1.29 & 0.95  \\
 $K^{(1)}_{\lambda_U}$   &  9.99 &  8.00 &  6.67 &  5.73 & 5.03 & 4.49 & 4.07 & 3.45 & 3.03 & 2.72 & 2.49  \\
\hline
 $K^{(0)}_{\lambda_B}$   & 25.27 & 14.93 & 9.57  &  6.51 & 4.63 & 3.42 & 2.60 & 1.61 & 1.08 & 0.77 & 0.57  \\
 $K^{(1)}_{\lambda_B}$   &  8.18 &  6.55 & 5.47  &  4.72 & 4.16 & 3.73 & 3.40 & 2.92 & 2.59 & 2.35 & 2.17  \\
\hline
 $K^{(0)}_{\lambda_V}$   & 14.93 &  8.65 & 5.46  &  3.67 & 2.60 & 1.91 & 1.45 & 0.90 & 0.61 & 0.43 & 0.32  \\
 $K^{(1)}_{\lambda_V}$   &  6.55 &  5.26 & 4.42  &  3.83 & 3.40 & 3.07 & 2.82 & 2.46 & 2.21 & 2.03 & 1.90  \\
\hline
 $K^{(0)}_{\lambda_R}$   &  8.27 &  4.70 & 2.93  &  1.96 & 1.38 & 1.02 & 0.78 & 0.49 & 0.33 & 0.24 & 0.18  \\
 $K^{(1)}_{\lambda_R}$   &  5.17 &  4.18 & 3.54  &  3.10 & 2.78 & 2.54 & 2.36 & 2.09 & 1.91 & 1.78 & 1.68  \\
\hline
\end{tabular}

\end{table}

\begin{table}
\caption[]{\small Various functions for bound-free transitions in negative hydrogen ions.}
\label{tab5}
\vspace{6mm}
\centering
\begin{tabular}{lllllllll}
\hline
$T_0$, $K $    &4$\cdot 10^3$ &5$\cdot 10^3$ &6$\cdot 10^3$ &7$\cdot 10^3$ &8$\cdot 10^3$ &9$\cdot 10^3$ &10$\cdot 10^3$ &12$\cdot 10^3$ \\
\hline
$K_{\lambda_U}$ &  3.21 & -1.92 & -4.08 & -4.98 & -5.31 & -5.36 & -5.28 & -4.96 \\
$K_{\lambda_B}$ & -4.79 & -6.29 & -6.57 & -6.43 & -6.12 & -5.79 & -5.46 & -4.87 \\
$K_{\lambda_V}$ & -9.14 & -8.40 & -7.57 & -6.83 & -6.20 & -5.67 & -5.24 & -4.56 \\
$K_{\lambda_R}$ &-10.73 & -8.84 & -7.51 & -6.54 & -5.81 & -5.25 & -4.81 & -4.16 \\
\hline
$\Delta(U-B)$   &-20.00 & -10.94& -6.24 & -3.62 & -2.04 & -1.06 & -0.45 &  0.21 \\
$\Delta(B-V)$   &-10.88 &  -5.26& -2.49 & -1.00 & -0.18 &  0.29 &  0.55 &  0.77 \\
$\Delta(V-R)$   & -3.97 &  -1.12&  0.14 &  0.71 &  0.96 &  1.05 &  1.07 &  1.01 \\
\hline

\end{tabular}
\end{table}

\end{document}